\documentstyle[prl,aps,epsf]{revtex}
\draft
\begin{document}
\twocolumn[\hsize\textwidth\columnwidth\hsize\csname @twocolumnfalse\endcsname

\title{Absolute Transverse Mobility and Ratchet Effect 
on Periodic 2D Symmetric Substrates}  
\author{C. Reichhardt and C.J. Olson Reichhardt}  
\address{ 
Center for Nonlinear Studies and Theoretical Division, 
Los Alamos National Laboratory, Los Alamos, New Mexico 87545}

\date{\today}
\maketitle
\begin{abstract}
We present a simple model of an overdamped particle
moving on a two dimensional symmetric periodic substrate with a dc drive
in the longitudinal direction and additional ac drives in
both the longitudinal and transverse directions.  
For certain regimes we find that a finite longitudinal dc force
produces a net dc response only in the transverse direction, which we
term absolute transverse mobility. Additionally we find regimes 
exhibiting a ratchet effect in the absence of an applied dc drive.
\end{abstract}

\vskip2pc]
\narrowtext
\section{Introduction}

When an overdamped particle is driven
with a dc drive, it moves in the direction of the drive, and
in the absence of any other external forces the particle 
velocity increases linearly with the drive. If there is some form of
pinning from a substrate, then in general for a finite
range of low drives the 
particle will be immobile or pinned \cite{Thorne1,Blatter2}. 
For higher drives in the
presence of pinning 
the velocity vs force curves can be highly nonlinear \cite{Thorne1,Blatter2}.  
For certain asymmetric substrate potentials, the particle
speed can decrease
with increasing applied drive, an effect that is termed 
negative differential
resistance \cite{Rachet3}. 
Such effects can also occur for
collections of classical coupled particles interacting with periodic
substrates \cite{Reichhardt4}.    
In addition a particle may
exhibit a finite average dc velocity in the {\it absence} of any external dc
drive.  This is often referred to as a ratchet effect,  
which can be thermal \cite{Ratchet5} or deterministic \cite{Det6}. 
Typically in ratchet systems
there is 
some form of underlying asymmetric potential 
which leads to a spatial symmetry breaking 
if the potential is flashed or if
an additional external ac drive is present.
In the case of {\it absolute negative mobility},  
when the particle is driven in the positive direction,
its motion is in the {\it opposite} (negative) direction.
Examples of this occur in
ratchet systems composed
of coupled particles, where the collective effects produce
the negative mobility \cite{Reimann7}. 
More recently a spatially  symmetric two-dimensional (2D) system  
was found which exhibits  absolute negative 
mobility for a {\it single} classical particle \cite{Hanggi8}.  
Many of these phenomena, such as negative differential resistance and
absolute negative mobility, 
also occur in 
various semiconductor
devices, where they arise due to quantum effects \cite{Scholl9}. 

In a 2D system, there are  
additional possibilities 
for the motion of an overdamped particle 
which are not available in 1D systems.  
Under an external dc drive in the longitudinal or $x$ direction, the
response can be a finite velocity in the $y$ or transverse direction only.
We call such a phenomenon {\it absolute transverse mobility}.  

In this work we present a simple model for a driven classical overdamped
particle moving in a 2D 
{\it symmetric} potential that exhibits a variety of 
dynamical behaviors, including phase locking, 
absolute transverse mobility, and 
ratchet effects.
Additionally we 
find
{\it reentrant} pinning phenomena where the moving particle becomes 
pinned upon increasing the drive.
Our system consists of a particle moving over a
symmetric periodic potential with an
applied dc drive in the $x$-direction and two additional ac drives
in both the $x$ and $y$ directions. We set
the amplitudes and frequencies for the two ac drives separately, and
also consider highly nonlinear combinations of the ac drives which 
produce asymmetric closed orbits. 
In all cases, in the absence of a substrate and dc drive
the average dc particle velocity is zero.  
Previous work on a similar system was performed with
much simpler circular particle orbits, produced by setting the amplitudes
of both ac drives equal to each other and fixing the
phase difference at $90^{\circ}$ 
\cite{Hastings10}. 

In the previous work,
several symmetrical phases were observed where the particle moves in 
both the transverse and longitudinal direction simultaneously when 
the dc drive is applied only in the longitudinal direction.
In the current work we consider the case where 
the amplitudes or frequencies of the two
ac drives are {\it different}. Here  a far 
richer variety of classical closed orbits are realizable.
The best examples of such orbits 
are Lissajous figures in which different sinusoidal 
orbits are plotted against one another.

Our results should apply to vortices moving
in superconductors 
with periodic pinning arrays or
Josephson junction arrays 
\cite{Moshchalkov11,Harada12,Look13,Schuller14,Kolton15} 
when ac currents are applied in both the transverse and longitudinal 
directions. Additionally, the behavior described here should be
observable in colloids moving over 2D periodic light arrays 
\cite{Korda16,Bechinger17} 
or through dynamically manipulated arrays of 
holographic tweezers \cite{Curtis18}. 
Further systems include biomolecules 
moving through periodic arrays of obstacles 
with two applied electric fields \cite{Viovy19}, electrons in a 
strong magnetic field moving through
2D antidot arrays with an additional ac drive to elongate one
direction of the electron orbit \cite{Weiss20}, or 
ions moving in dissipative optical
trap arrays \cite{Grynberg21} with ac applied fields.  
In the
case of superconductors our results can also have applications
for the controlled motion or removal of flux from superconductors and SQUIDs.
For colloids and biomolecules our results could 
provide a useful method for separating
different particle species.  

\section{Model}

In our model 
we consider an overdamped particle 
moving 
over a 2D periodic substrate. 
The  equation of motion
is:
\begin{equation}
{\bf f} =  
{\bf f}_{s} + {\bf f}_{DC} + {\bf f}_{AC} = 
\eta \frac{d{\bf r}}{dt} 
\end{equation} 
with the damping constant
$\eta=1$. 
The assumption of overdamped motion
should be valid for vortices in superconductors or
Josephson junction arrays as well as for colloidal systems.  
The substrate consists of a periodic square array of obstacles with
lattice constant $a$. 
The force from the substrate, composed of a fixed square array
of repulsive particles,  
is ${\bf f}_{s} = \sum -\nabla U(x,y)$.
This substrate can be realized in 
superconductors with a square periodic array of holes
\cite{Moshchalkov11,Harada12} or
magnetic dots \cite{Schuller14} when each site captures one vortex and
additional vortices sit in the interstitial regions between the sites.
The interstitial vortices move in a periodic potential
created by the repulsive interaction from the pinned vortices
so that $U(r) = F_0\ln(r)$ for 
$r\ll 2\lambda^2/d$.
Here forces are measured in units of
$F_0 = \Phi_0^2 d/16\pi\lambda^2$,
where $\Phi_0$ is
the flux quantum, $d$ is the film thickness, and $\lambda$ is the London
penetration depth.  This potential can be treated as in Ref.~\cite{Gronbech}.
The motion of interstitial vortices has been directly imaged 
in experiments with this geometry \cite{Harada12}, and
phase-locking effects, dc, and ac driven interstitial vortex motion 
have also been observed  \cite{Look13}.
For most of the results presented here we use
a system of size $8a \times 8a$. For larger systems we
observe the same results, indicating that our system is large enough
to capture the essential physics. 
We have tested
different initial conditions by placing the particle at different
locations at the start of the simulations and find that the
general results are unchanged.
Additionally, we have considered other potentials, such as those created by 
fixed Coulomb charges or Yukawa potentials, and find that they produce the 
same phases.  

Throughout this work the 
dc drive  ${\bf f}_{DC}$=$f_{DC}{\hat {\bf x}}$ is applied
along the positive longitudinal ($x$ direction) symmetry axis of the  
pinning array.
The ac drive is applied in both the $x$ and $y$ directions and  
is given by
\begin{equation}
{\bf f}_{AC} = A\sin(\omega_{A}t){\hat {\bf x}} - 
B\cos(\omega_{B}t){\hat {\bf y}}.
\end{equation}

\begin{figure}
\center{
\epsfxsize=3.5in
\epsfbox{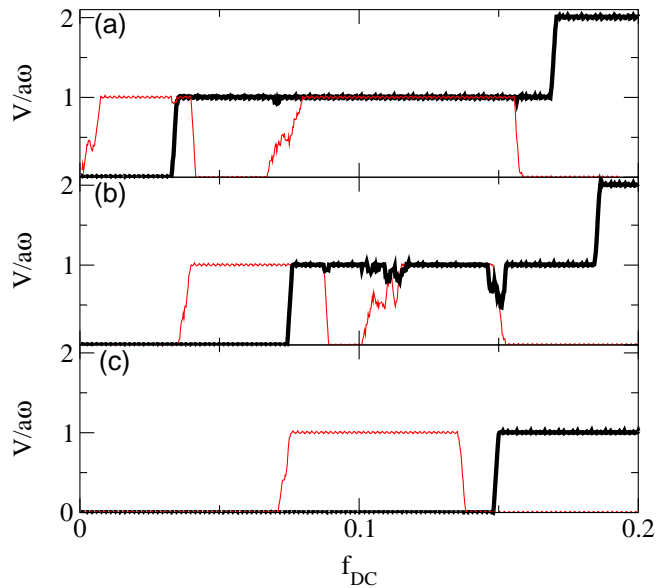}}
\caption{The longitudinal particle velocity 
$V_{x}/a\omega$ (heavy lines) and 
transverse particle velocity $V_{y}/a\omega$ (light lines) vs applied dc drive
${\bf f_{DC}}=f_{DC}{\hat {\bf x}}$ 
for ac amplitudes (a) $A/B = 0.875$, (b) $A/B = 0.625$,
and (c) $A/B = 0.375$. 
}
\end{figure}

\noindent
Note that there is {\it no} 
dc driving component in the transverse or $y$ direction. 
In the first part of this work we set $\omega_A = \omega_B$,
fix $B = 0.24$, and vary $A$.
For $A = B$ the particle moves in a clockwise circle
just large enough to encircle one maximum of the substrate
potential. 
As $A$ decreases the particle orbit becomes elliptical with the 
long side in the $y$-direction.
We monitor the time averaged particle velocity 
$V_{x} = \frac{1}{TN}\sum_{t=0}^{T}\sum_{i=0}^{N}v_{i}(t) \cdot {\bf \hat{x}}$,
where $T$ is the period and $N$ is the number of particles,
and similarly the time averaged transverse velocity $V_{y}$ as
${\bf f}_{DC}$ is increased from 0 to $1.0$
in increments of $2.5 \times 10^{-4}$,
with $3 \times 10^5$ time steps spent at each drive
to ensure a steady state. 
Time is measured in units of $t_0=\eta/F_0$.
We also investigate 
$\omega_{B} \neq \omega_{A}$ for varied ac amplitude. In this case we 
find a ratchet effect.   

\section{Absolute Transverse Mobility} 

In Fig.~1 we plot $V_{x}$ 
(heavy line) and $V_{y}$ (light line) vs $f_{DC}$ for different values of
$A/B$ at fixed $\omega_A/\omega_B = 1$ and $a=1.42 \lambda$. 
In Fig.~1(a) at $A/B = 0.875$, 
$V_{x} = 0$ for $f_{DC} < 0.03$, indicating that the particle
is pinned in the $x$-direction. For $f_{DC} \ge 0.03$, 
$V_{x}$ increases in a series of steps of height $a\omega$. 
These steps are a signature of the 
phase locking which occurs due to resonances between
the applied ac frequency and the washboard frequency generated as the
particle moves over the periodic substrate. 
As shown in Fig.~1(a), $V_{y}$ has
a finite value of $V_{y}=a\omega$
for $f_{DC} < 0.03$, indicating that even though the dc drive is
strictly in the $x$-direction the particle is {\it moving strictly in the
$y$-direction}. In analogy with the 

\begin{figure}
\center{
\epsfxsize=3.5in
\epsfbox{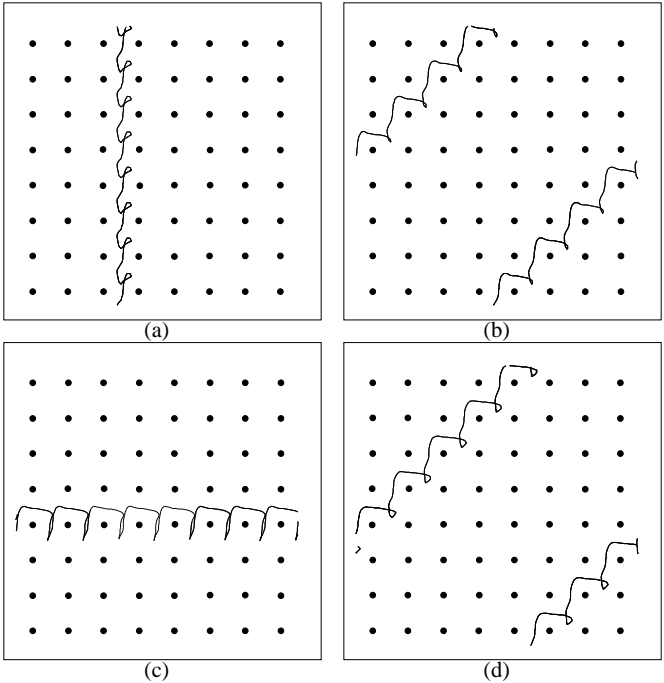}}
\caption{The particle trajectories (black lines) for 
a fixed time interval for the system in Fig.~1(a) with $A/B=0.875$. 
The black dots are the potential maxima from 
the periodic substrate. (a) $f_{DC} = 0.025$, Phase II$_{y}$; 
(b) $f_{DC} = 0.0375$,
Phase III$_{x-y}$; 
(c) $f_{DC} = 0.06$, Phase IV$_x$; and (d) $f_{DC} = 0.125$,
Phase III$_{x-y}$.    
}
\end{figure}

\noindent
phenomena of absolute negative mobility, 
where a particle moves in the opposite direction of an applied driving
force, we term the strictly $y$ direction motion 
{\it absolute transverse mobility}. 
For large enough drives, $f_{DC} \ge 0.16$, the motion is strictly in the 
$x$ direction. 
At intermediate drives $0.03 < f_{DC} < 0.16$,
different dynamical phases appear. For
$0.045 < f_{DC} < 0.065$ the particle moves in the $x$-direction only, 
while for
$0.07 < f_{DC} < 0.158$, $V_{x}=V_{y}$, indicating
that the particle is moving at $45^{\circ}$ with respect to the drive. 
There is also a small region near $f_{DC} = 0.035$ where $45^{\circ}$ 
motion occurs. 
In Fig.~1(b) for $A/B = 0.625$ there is a clear {\it pinned} 
phase at low $f_{DC}$ where both $V_{x}$ and $V_{y}$ are zero. 
As we increase the drive, we observe the same phases shown 
in Fig.~1(a), with the boundaries shifted. 
Near the transitions of these phases, smaller steps
in $V_{x}$ and $V_{y}$ can occur with height $pa\omega/q$
where $p$ and $q$ are integers. In Fig.~1(c), for $A/B = 0.375$, 
$45^{\circ}$ motion no longer appears. Instead, there is a
remarkable {\it reentrant} pinned phase for $0.1385 < f_{DC} < 0.148$. 
As $f_{DC}$ increases, the particle is first pinned, then moves in the
$y$-direction, is repinned again, and finally moves in the $x$-direction.  

We next consider the particle trajectories in these different phases. 
We term the pinned regime Phase I$_p$, the absolute transverse mobility regime
Phase II$_y$, the 
$45^{\circ}$ motion regime Phase III$_{x-y}$, and 
the strictly $x$ direction motion Phase IV$_x$. In Fig.~2 we illustrate
these phases 

\begin{figure}
\center{
\epsfxsize=3.5in
\epsfbox{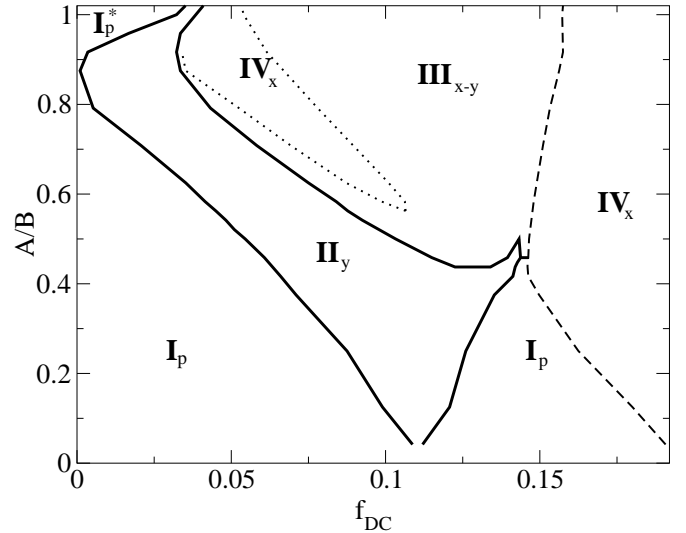}}
\caption{Dynamic phase diagram for $A/B$ vs $f_{DC}$
for a system with $\omega_A=\omega_B$.
Phase I$_p$: pinned phase.  Phase II$_{y}$: motion only in the $y$-direction. 
Phase III$_{x-y}$: motion at $45^\circ$. Phase IV$_x$:
motion only in the $x$-direction. }
\end{figure}

\noindent
for fixed values of $f_{DC}$
from the system in Fig.~1(a) with $A/B = 0.875$.
The black lines are the
trajectories of the moving particle and the black dots are the 
potential maxima of the underlying periodic substrate. 
Figure 2(a) shows the trajectory of the moving particle 
in Phase II$_y$ at $f_{DC}=0.025$. 
In every period the particle makes a small loop, but 
the net motion of the particle is in the positive $y$ direction only. 
In Phase III$_{x-y}$, shown in Fig.~2(b) for $f_{DC}=0.0375$, the particle 
moves equal distances in the $x$ and $y$ directions.
At $f_{DC} = 0.06$, illustrated
in Fig.~2(c), the Phase IV$_x$ motion is strictly in 
the $x$ direction, and the particle translates a distance $a$ every period.
In Fig.~2(d) we show the reentrant Phase III$_{x-y}$ flow for $f_{DC} = 0.125$,
very similar to that seen in Fig.~2(b). For higher drives
$f_{DC} > 0.16$, the motion is strictly in the $x$ direction and 
is similar to 
Fig.~2(c). 

For the simulations shown in Figs.~1 and 2,
the particle initially begins in the 
center of the plaquette with zero dc drive.
We considered the effect both of using different starting locations, 
and of setting $f_{DC}$ to a finite initial value.
In each case we find that the particle quickly
settles into a regular orbit. The orbits that 
appear when $V_{x}$ or $V_{y}$ are
constant correspond to the same periodic, phase locked attractor orbits
that were obtained previously in Figs. 1 and 2. 
We do 
find some non-periodic orbits in the regions where $V_{x}$ or $V_{y}$
are not constant, 
such as in the transition regimes between the different phases.

We performed a series of simulations at different values of $A/B$
to identify the onset of the four 
phases as a function of $f_{DC}$. 
In Fig.~3 we present the resulting dynamic phase diagram $A/B$ vs $f_{DC}$
which shows a very rich structure.
For $A/B = 0$ the system depins directly into Phase IV$_x$ and there are no phases 
that involve motion in the $y$-direction.
Phase II$_y$ first occurs for $A/B>0.03$, and gradually increases in width
until 
$A/B \approx 0.43$, when Phase III$_{x-y}$ appears.
The reentrant pinning Phase I$_p$ decreases in size 
and then disappears over this same interval.
There is also a small region around $A/B \gtrsim 0.43$
where Phase II$_y$ and Phase III$_{x-y}$ are both 
reentrant.
A reentrant tongue of Phase IV$_x$ at low drives
appears for $A/B > 0.55$. At all values of $A/B$, the
flow at large $f_{DC}$ is strictly in the $x$ direction 
(Phase IV$_x$).  The width of the pinned Phase I$^*_p$ increases upon
approaching $A/B = 1$ from below
since the almost circular particle orbit
around one potential maximum is highly stable. 
Motion in the $y$ direction (Phase II$_y$) still appears
for the symmetric ac drive $A/B=1$ 
due to the fact that the particular chirality of the ac drive
breaks the reflection symmetry across the $y$ axis.

For $A/B > 0.6$, the sliver of reentrant Phase
 III$_{x-y}$ falling between Phases II$_y$ and IV$_x$ 
becomes gradually smaller as $A/B$ increases 
until it vanishes above $A/B=0.88$.
The transition from Phase II$_y$ to Phase IV$_x$ 
at higher values of $A/B>0.88$
occurs in a very small window of
of $f_{DC}$ but is not completely sharp. Instead there is a small region
where the flowing particle moves in both the $x$ and $y$ direction but at 
an angle less than $45^\circ$. The flow is intermittent and jumps
among different orbits or angles. We have also considered 
the effect of starting the particle at a nonzero fixed $f_{DC}$ 
in the transition region between Phases II$_y$ and IV$_x$
and find that the flow will settle quickly to
Phase II$_y$ or IV$_x$.  

We now consider the conditions under which
transverse mobility and the reentrant pinning can occur, and indicate 
where the boundaries between the different phases are expected to fall.

{\it Pinned phases I$_p$ and I$^*_p$}: The particle remains in the pinned
phase I$_p$ as long as the combined dc and maximum ac components are less
than the confining barrier produced by the repulsive obstacles.  This
barrier has a strong angular dependence due to its egg-carton shape, 
and the lowest points of the barrier fall at the center of each of the
four sides of the plaquette, at $x_{\rm min}^{\pm}$ and $y_{\rm min}^{\pm}$,
where the x- or y-confining forces pass through a minimum.  The largest thresholds
occur for a particle trajectory along a 45$^\circ$ angle passing through
the potential maximum.  At $A/B \approx 0.88$, a transition to a new pinned
phase I$_p^*$
occurs.  For $A/B<0.88$, the pinned particle orbit is contained inside
a single plaquette.  For $A/B>0.88$, the orbit becomes too large to fit inside
the plaquette, and the particle switches to a larger orbit centered around
one of the potential maxima.

{\it Transition from I$_p$ to II$_{y}$}:
For $A=0$ and fixed B, the particle orbit consists of a single line extending from
the top to the bottom of the plaquette, close to the minimum y-confining force
points of the potential, $y_{\rm min}^{\pm}$.  Due to this proximity induced
by the y-component of the ac drive, the particle depins in the y-direction
{\it before} it depins in the x-direction for nonzero values of A, and enters
phase II$_{y}$.  The particle hops from one plaquette to the next plaquette in
the positive y-direction during a brief interval at the end of the rising phase
of the particle orbit.  In order to make this hop, the particle must be moving
rapidly enough to reach the next plaquette before the ac phase reaches the
downward portion of the cycle.  Increasing $f_{DC}$ or increasing $A$ both
contribute to increasing the velocity of the particle during the hop.  The
minimum $f_{DC}$ value required to permit the particle to hop at low but
nonzero $A$ is $f_{DC}\approx 0.11$.  As $A/B$ increases, the x-component of
the ac force also contributes to the particle velocity during the hop, so
the value of $f_{DC}$ that must be applied to induce y-direction motion drops,
such that the particle velocity during the hop at the onset of phase II$_{y}$
remains roughly constant as $A/B$ increases.

{\it Transition from II$_{y}$ to I$_p$}:
As $f_d$ increases further within Phase II$_{y}$, for $A/B<0.45$, a transition
to a {\it reentrant} pinned phase occurs.  Here, the x-velocity of the particle
has increased enough that the particle is swept past the minimum in the
y-confining potential, $y_{\rm min}^{+}$, before it has time to hop to the
next plaquette.  Thus the particle returns to a pinned orbit.  The II$_{y}$-I$_p$
transition line moves to higher $f_d$ with increasing $A/B$ due to the fact that
the time required for the particle to complete its hop to the next plaquette
drops as $A/B$ increases.  Therefore, a higher $f_{DC}$ is required to 
sweep the particle past $y_{\rm min}^{+}$ before the hop is complete.

{\it Transition from I$_p$ to IV$_x$}:
Beyond the second pinned phase I$_p$, as $f_{DC}$ is further increased,
the particle orbit is shifted closer to the minimum in the x-confining
potential on the right side of the plaquette, $x_{\rm min}^{+}$,
and when the combined ac and dc forces in the x-direction exceed the
potential strength at this minimum, the particle depins in the positive
x-direction.  For $A/B < 0.45$, $f_{DC}$ at the depinning transition
decreases with increasing $A/B$ due to the fact that the ac force
also contributes to the net x-force on the particle.  This 
contribution saturates at $A/B \approx 0.45$, when the
II$_y$-I$_p$ line meets the I$_p$-IV$_x$ line and
the reentrant phase disappears.

{\it Transition from II$_y$ to IV$_x$, with a reentrant III$_{x-y}$}:
For $A/B>0.45$, as $f_{DC}$ is increased, the particle motion 
leaves Phase II$_{y}$, passing briefly through a sliver of
Phase III$_{x-y}$ before entering phase IV$_x$, with
motion in the x-direction only.  Depinning of the particle in the
x-direction first occurs when the x-component of the ac drive
combined with $f_{DC}$ exceeds the x-confining force of the
potential at $x_{\rm min}^{+}$.  At this value of $f_{DC}$,
the particle is moving in both the x and y directions in phase
III$_{x-y}$.  Due to the x-motion of the particle, however, the
y-motion becomes unstable, and at slightly higher values of
$f_{DC}$, the y-motion ends and the particle moves only in
the x-direction in phase IV$_x$.  As the drive is further increased,
stable y-motion becomes possible again, and the particle enters
a wide region of phase III$_{x-y}$.

{\it Transition from III$_{x-y}$ to IV$_x$ at high $f_{DC}$}:
As $f_{DC}$ is increased further, the y-direction motion of
phase III$_{x-y}$ 

\begin{figure}
\center{
\epsfxsize=3.5in
\epsfbox{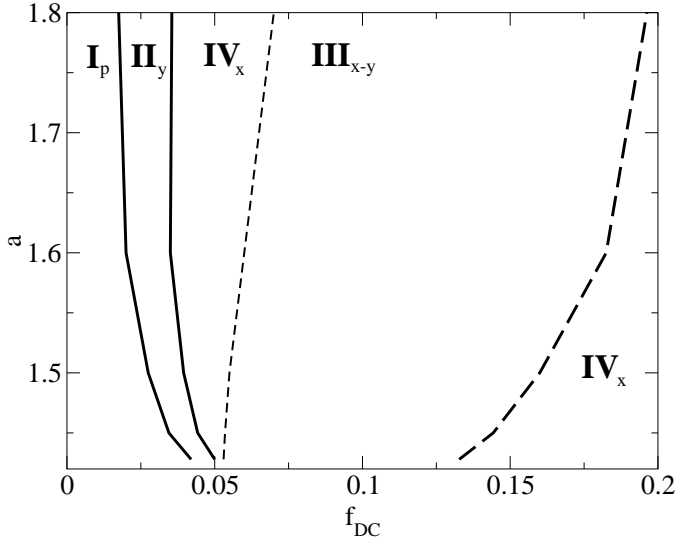}}
\caption{Dynamic phase diagram for 
substrate lattice constant $a$ vs $f_{DC}$
for a system with $A = B$ and $\omega_A=\omega_B$.
Phase I$_p$: pinned phase.  Phase II$_{y}$: motion only in the $y$-direction. 
Phase III$_{x-y}$: motion at $45^\circ$. Phase IV$_x$:
motion only the in $x$-direction. }
\end{figure}

\noindent
ends when the particle orbit becomes so
extended in the x-direction due to the increased $f_{DC}$ that
it can no longer depin in the y-direction since it has moved
away from the y-minimum location $y_{\rm min}^{+}$.

We next consider the effect of changing the density of the system. 
We achieve this by performing
a series of simulations for fixed $A/B = 1.0$
while varying the lattice constant of the periodic substrate
and changing the system size accordingly. 
This increases
the effective substrate strength since it increases
the barrier to hop from one plaquette to
another.
In Fig.~4 we show the phase diagram 
of the substrate lattice constant $a$ vs $f_{DC}$. 
We consider lattice constants ranging from $1.42\lambda$ to $1.8\lambda$.  
New dynamic phases appear for lattice constants
outside of this range. For smaller
lattice constants $a<1.42$, the zero-dc drive orbits start to encircle two or more 
potential maxima. For larger lattice constants, 
$a>1.8$, the phases start to show 
disordered or chaotic behavior, and Phases I$_p$ to 
IV$_x$ become difficult to 
define. Fig.~4 shows that the width of 
the pinned Phase I$_p$ grows for denser systems,
as expected due to the increased barriers for inter-plaquette jumps.
It might be expected that Phase II$_y$ would grow for smaller $a$ as the
plaquettes shrink; however, the increased repulsion 
caused by the shorter distance between the potential maxima
and the particle causes the particle orbit to shrink as well.
Both Phase II$_y$ and the first Phase IV$_x$ shrink
with decreasing $a$, while the second onset of 
Phase IV$_x$ occurs at a lower value of $f_{DC}$
since the particle does not need to move as far in
the x-direction in order to reach the next plaquette. 

We have also considered the effects of adding a phase shift to the ac drives.
We shift the phase of the x-component of the
ac drive by $\delta$: 
$A\sin(\omega_{A} t + \delta){\bf {\hat x}}$. 
This 

\begin{figure}
\center{
\epsfxsize=3.5in
\epsfbox{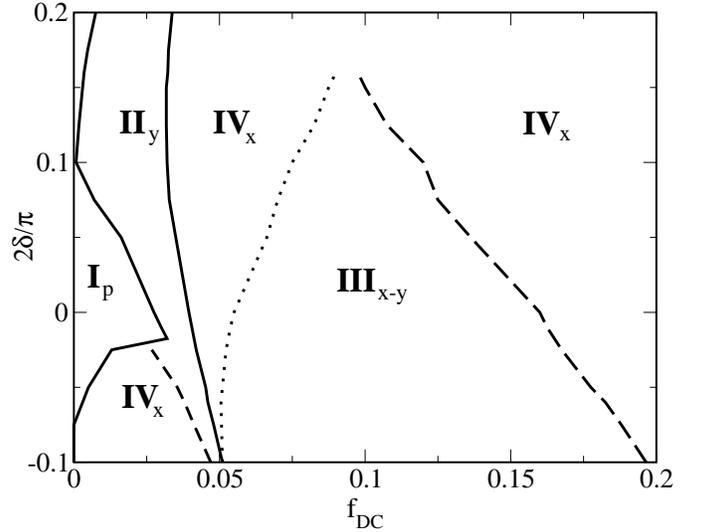}}
\caption{Dynamic phase diagram for the phase 
shift $\delta$ in units of $2/\pi$ vs $f_{DC}$
for a system with $\omega_A=\omega_B$ and $A = B$.
The phase shift is added to the $x$-component of the ac drive. 
Phase I$_p$: pinned phase.  Phase II$_{y}$: motion only in the $y$-direction. 
Phase III$_{x-y}$: motion at $45^\circ$. Phase IV$_x$:
motion only the in $x$-direction. }
\end{figure}

\noindent
causes the particle orbits
to become tilted and elliptical. 
We restrict ourselves to the region $-0.05/\pi < \delta < 0.1/\pi$ 
in phase space where the
four phases described above occur. For larger shifts new phases can 
arise. 
In Fig.~5 we plot the phase diagram as a function of
phase shift $\delta$ vs $f_{DC}$. 
At $\delta=\pi/2$ the ac orbit would be a straight line along $45^\circ$.    
Fig.~5 shows that as $\delta$ increases from zero, which corresponds to the 
elliptical orbit being tilted toward the right, 
Phase II$_{y}$ increases in size and
Phase I$_p$ decreases in size. 
For large enough shifts $\delta$,
Phase III$_{x-y}$ disappears. For 
negative phase shifts 
a new region of Phase IV$_x$ appears between 
Phases I$_p$ and II$_{y}$. 
For large enough negative phase shifts
 we observe a ratchet effect in which the 
particle moves in the $x$-direction with zero dc drive. We discuss this
more in the next section. 
Also for increasing negative phase shifts, 
Phase II$_{y}$ decreases in size while
Phase III$_{x-y}$ increases in size.  

{\section{Ratchet Effects}

The elliptical ac drives in the previous sections 
preserve the combined $x$ and $y$ 
reflection symmetries of the system.
ac drives with additional asymmetries 
can produce a net dc motion or ratchet effect in the absence of 
a dc drive. 
We consider a system with $f_{DC} = 0$ under
an asymmetric applied ac drive of the form 
\begin{equation}
{\bf f}_{AC} = 
A[\sin(\omega_{A}t) + \sin^{3}(\omega_{B}t)]{\bf {\hat x}} -
B\cos(\omega_{B}t){\bf {\hat y}}
\end{equation} 
 where 
$\omega_B/\omega_A = 0.8$.
In Fig.~6(a) 
we plot $V_{y}$ vs $A$ for constant $B=0.34$
and $f_{DC}=0$. In this regime, $V_{x} = 0$.
There are a series of regions for increasing 
$A$ that have 

\begin{figure}
\center{
\epsfxsize=3.5in
\epsfbox{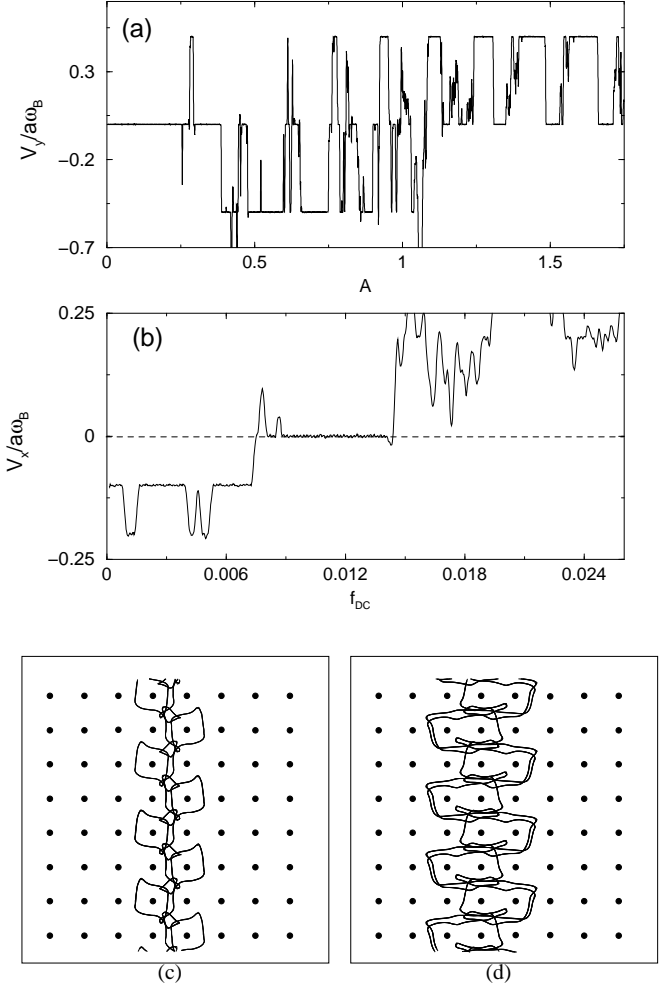}}
\caption{
(a) Transverse velocity $V_{y}/a\omega_B$ 
vs $A$ at $f_{DC} = 0$ and $B=0.34$ for a system with an asymmetric 
ac drive, $\omega_B/\omega_A=0.8$. 
(b) Longitudinal velocity 
$V_{x}$ vs $f_{DC}$ for a system at $B/A=0.7$
showing a negative ratchet effect 
at $f_{DC}<0.0075$. 
(c) Particle trajectory for a positive $V_y$ orbit from panel (a) at
$A=0.285$, $B=0.34$, and $f_{DC}=0$.
(d) Particle trajectory for a negative $V_y$ orbit from panel (a) at 
$A=0.5$, $B=0.34$, and $f_{DC}=0$.}
\end{figure}

\noindent
a finite dc value of $V_{y}$ in both the
positive and negative direction, indicating a ratchet effect with 
velocity $a\omega_{B}/2$. 
In Fig.~6(c) we illustrate a positive $V_{y}$ orbit  
from the system in Fig.~6(a) at $A = 0.285$.
The orbit has alternating lobes which are angled in the positive $y$ 
direction. 
In Fig.~6(d) we show that the negative $V_{y}$ orbit at $A = 0.5$ 
is composed of an alternating double rectangular orbit which is tilted in the
negative $y$ direction.  

We have also found ratchet regimes where there is a finite dc velocity in the
{\it opposite} direction of the applied dc drive. 
In Fig.~6(b), we plot $V_{x}$ vs $f_{DC}$ for a system 
with ${\bf f}_{AC} = A\sin(\omega_{A}t){\bf {\hat x}} - 
B\cos(\omega_{B}t){\hat {\bf y}}$, $B/A = 0.7$, and 
$\omega_{A}/\omega_{B} = 1.85$. 
Here there is a regime $0 < f_{DC} < 0.0075$ 
where the particle is moving 
{\it backward} with respect to 
the dc drive, which is applied in the positive $x$-direction. 

\begin{figure}
\center{
\epsfxsize=3.5in
\epsfbox{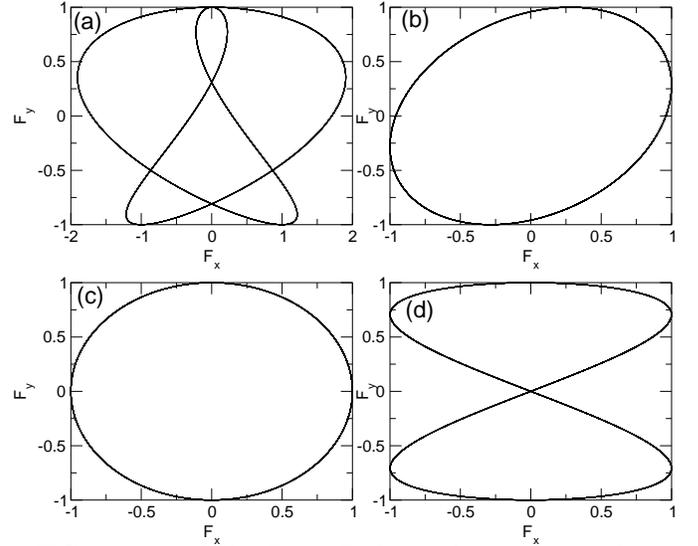}}
\caption{(a) and (b) $F_{x}$ vs $F_{y}$ for ac drives 
that produce a ratchet effect. (a) 
${\bf f}_{AC} = A\sin(\omega_{A}t){\bf {\hat x}} + 
A\sin(1.5\omega_{A}t){\bf {\hat x}} -  B\cos(1.5\omega_{B}t){\bf {\hat y}}$,
$A/B=1$, $A=1$, and $\omega_A/\omega_B=1$.
(b) A ratchet effect produced by a phase shift:
${\bf f}_{AC} = A\sin(\omega_{A} t + \delta){\bf {\hat x}}
- B\cos(\omega_{B}t)$,
$\delta=0.287$, $A/B=1$, and $\omega_A/\omega_B=1$.
(c) and (d) $F_{x}$ vs $F_{y}$ for 
ac drives that do not produce a ratchet effect.
(c) ${\bf f}_{AC} = A\sin(\omega_{A}t){\bf {\hat x}} - B\cos(\omega_{B}t)
{\bf {\hat y}}$ with $A/B=1$ and $\omega_A/\omega_B=1$.
(d) ${\bf f}_{AC} = A\sin(\omega_{A}t){\bf {\hat x}} -
 B\cos(2\omega_{B}t){\bf {\hat y}}$ with $A/B=1$ and $\omega_A/\omega_B=1$. 
}
\end{figure}

\noindent
The system enters a pinned phase for
$0.009 < f_{DC} < 0.014$ before beginning to move strictly in the
positive $x$-direction for $f_{DC} > 0.014$.
We note that this negative velocity in opposition to the dc drive
is not a negative mobility regime. Instead it is a ratchet effect
which can persist for a range of opposite dc drive. The fact that 
there is a finite negative velocity even at $f_{DC} = 0.0$ as seen in 
Fig.~6(b) shows that the dc drive is not causing the net dc motion. 
 
A basic question is what are the minimal ac drive criteria required to
produce a ratchet effect at $f_{DC} = 0.0$.  
In general we find that ratchet effects occur
for ac drives in which at least one of the spatial reflection symmetries  
is broken. 
An example of a simple ac drive that produces a zero-dc
ratchet effect is 
${\bf f}_{AC} = A\sin(\omega_{A}t){\bf {\hat x}} + 
A\sin(1.5\omega_{A}t){\bf {\hat x}} -  
B\cos(1.5\omega_{B}t){\bf {\hat y}}$,
with $A/B=1$ and $\omega_A/\omega_B=1$.   
In Fig.~7(a) we plot 
$F_{x}$ vs $F_{y}$  for this ac drive
in the absence of a substrate with $A = 1.0$.  
In Fig.~8(a) we show the particle motion over the substrate for
$f_{DC} = 0.0$.  The particle translates in the negative $y$-direction.     
Fig.~7(a) shows that the ac drive breaks a spatial symmetry across the
$y$-axis. The orbit of the particle thus breaks an $x$-axis symmetry. 
In Fig.~7(c) 
we plot $F_{x} $ vs $F_{y}$ for 
${\bf f}_{AC} = A\sin(\omega_{A}t){\bf {\hat x}} - B\cos(\omega_{B}t)
{\bf {\hat y}}$ with $A/B=1$ and $\omega_A/\omega_B=1$,
and in Fig.~7(d) we plot $F_{x}$ vs $F_{y}$ for  
${\bf f}_{AC} = A\sin(\omega_{A}t){\bf {\hat x}} -
 B\cos(2\omega_{B}t){\bf {\hat y}}$ with $A/B=1$ and $\omega_A/\omega_B=1$. 
These drives
do not produce a zero-dc ratchet effect, and these
orbits do not break 

\begin{figure}
\center{
\epsfxsize=3.5in
\epsfbox{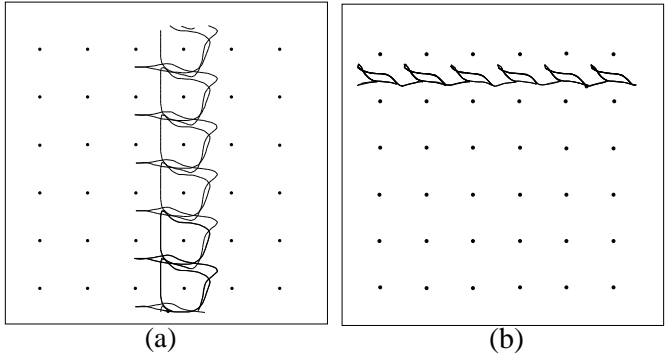}}
\caption{
The particle trajectories 
(black line) for a fixed time interval
and $f_{DC} = 0.0$
for motion in a 2D periodic substrate with potential maxima
located at the black dots. 
(a) A ratchet effect due to the breaking of a reflection symmetry.
The particle moves in the negative y-direction when 
driven with the ac drive shown in 
Fig.~7(a), 
${\bf f}_{AC} = A\sin(\omega_{A}t){\bf {\hat x}} + 
A\sin(1.5\omega_{A}t){\bf {\hat x}} -  B\cos(1.5\omega_{B}t){\bf {\hat y}}$,
$A/B=1$, $\omega_A/\omega_B=1$.
(b) A ratchet effect produced by the addition of a phase shift.
The particle moves in the positive x-direction when 
driven with the ac drive shown in Fig.~7(b),
${\bf f}_{AC} = A\sin(\omega_{A} t + \delta){\bf {\hat x}}
- B\cos(\omega_{B}t)$,
$\delta=0.287$, $A/B=1$, and $\omega_A/\omega_B=1$. 
}
\end{figure}

\noindent
a reflection symmetry.  
As indicated in the phase diagram of Fig.~6,
the addition of a phase shift to the ac drive in the
x-direction produces
another simple ac drive that exhibits a zero-dc ratchet effect. 
In Fig.~8(b) we show a particle orbit under the addition
of a negative phase shift, 
${\bf f}_{AC} = A\sin(\omega_{A} t + \delta){\bf {\hat x}}
- B\cos(\omega_{B}t)$
with $\delta=0.287$, $A/B=1$, and $\omega_A/\omega_B=1$,
that produces a zero-dc ratchet effect
in the x-direction.
$F_x$ vs. $F_{y}$ for this orbit is illustrated in Fig.~7(b).
The ac drive in Fig.~7(b) breaks both the $x$ and $y$ 
reflection symmetries.    
We note that even if an ac drive breaks a reflection symmetry
it does not necessarily produce a zero-dc ratchet. However, 
we have never observed 
zero-dc ratcheting for symmetrical ac drives. 
 
\section{Summary}

We have presented a simple model 
of a particle driven in a two-dimensional 
symmetric periodic substrate
with an elliptical ac drive. 
In certain regimes, the particle moves strictly in the 
$y$-direction when the dc drive is applied in the $x$-direction. 
We term this effect absolute transverse mobility. 
Additionally we have mapped several dynamic phase diagrams for this system,
which feature a number of dynamical phases including simultaneous motion 
in the $x$ and $y$ direction, as well as a reentrant pinned phase. 
For ac drives which produce asymmetric orbits,
net dc motion or a zero-dc ratchet effect
can arise in the absence of an applied dc drive. 
Our results can be tested experimentally and have
useful applications for controlling flux motion
in superconductors, colloids in optical trap arrays,
and biomolecules moving through a periodic obstacle array.  

We thank C. Bechinger, D. Grier, M.B. Hastings, and
P. Korda for useful discussions.  
This work was supported by the US DOE under Contract No. W-7405-ENG-36.


\begin{references}

\bibitem{Thorne1}
G.~Gr{\" u}ner, Rev.~Mod.~Phys.~{\bf 60}, 1129 (1988).

\bibitem{Blatter2}
G.~Blatter, M.V. Feigel'man, V.B. Geshkenbein, A.I. Larkin,
and V.M. Vinokur, Rev.~Mod.~Phys.~{\bf 66}, 1125 (1994).

\bibitem{Rachet3}
G.A. Cecchi and M.O.~Magnasco, Phys.~Rev.~Lett.~{\bf 76}, 1968 (1996). 

\bibitem{Reichhardt4}
C.~Reichhardt, C.J.~Olson and F.~Nori, Phys.~Rev.~B {\bf 58}, 6534 (998).

\bibitem{Ratchet5}
M.O.~Magnasco, Phys.~Rev.~Lett.~{\bf 71}, 1477 (1993);
R.D.~Astumian and M.~Bier, 
{\it ibid.}~{\bf 72}, 1766 (1994);
C.R.~Doering, W.~Horsthemke, and J.~Riordan, 
{\it ibid.}~{\bf 72}, 2984
(1994); P. ~Reimann, Phys.~Rep.~{\bf 361}, 57 (2002). 

\bibitem{Det6}
R.~Bartussek, P.~H{\" a}nggi, and J.C. Kissner, Europhys. Lett.~{\bf 28}, 459
(1994); 
J.L.~Mateos, Phys.~Rev.~Lett.~{\bf 84}, 258 (2000). 

\bibitem{Reimann7}
P.~Reimann, R. Kawai, C. van den Broeck, and P. H{\" a}nggi, 
Europhys.~Lett.~{\bf 45}, 545 (1999);
J.~Buceta, J.M.~Parrondo, C. Van den Broeck, and F.J.~de la Rubia,
Phys.~Rev.~E {\bf 61}, 6287 (2000);
B.~Cleuren and C. Van den Broeck, 
Europhys.~Lett.~{\bf 54}, 1 (2001).

\bibitem{Hanggi8}
R.~Eichhorn, P.~Reimann, and P.~H{\" a}nggi, Phys.~Rev.~Lett.~{\bf 88}, 190601 
(2002); Phys.~Rev.~E.~{\bf 66}, 066132 (2002).

\bibitem{Scholl9} 
K.~Seeger, {\it Semiconductor Physics} (Springer-Verlag, Berlin, 1982);
E.~Scholl, {\it Nonequilibrium Phase Transitions in Semiconductors}
(Springer-Verlag, Berlin, 1987); S.~Wang and D.D.L.~Chung,
Composites B, {\bf 30}, 579 (1999).

\bibitem{Hastings10}
C. Reichhardt, C.J. Olson and M.B. Hastings, Phys.~Rev.~Lett.~{\bf 89},
024101 (2002). 

\bibitem{Moshchalkov11}
M.~Baert, V.V. Metlushko, R. Jonckheere, V.V. Moshchalkov,
and Y. Bruynseraede, Phys.~Rev.~Lett.~{\bf 74}, 3269 (1995);
S.B.~Field, S.S. James, J. Barentine, V. Metlushko, G. Crabtree,
H. Shtrikman, B. Ilic, and S.R.J. Brueck, 
Phys.~Rev.~Lett.~{\bf 88}, 067003 (2002). 

\bibitem{Harada12}
K.~Harada, O. Kamimura, H. Kasai, T. Matsuda, A. Tonomura,
and V.V. Moshchalkov, Science {\bf 274}, 1167 (1996).
   
\bibitem{Look13}
L.~Van Look, E. Rosseel, M.J. Van Bael, K. Temst, V.V. Moshchalkov,
and Y. Bruynseraede, Phys.~Rev.~B {\bf 60}, R6998 (1999).

\bibitem{Schuller14} 
J.I.~Mart{\' \i}n, M. V{\' e}lez, A. Hoffmann, I.K. Schuller,
and J.L. Vicent, Phys.~Rev.~Lett.~{\bf 83}, 1022 (1999).

\bibitem{Kolton15}
C.~Reichhardt, A.B. Kolton, D. Dom{\' \i}nguez, and N. Gr{\o}nbech-Jensen,
Phys.~Rev.~B {\bf 64}, 134508 (2001).

\bibitem{Korda16}
P.T. Korda, M.B. Taylor, and D.G. Grier, Phys.~Rev.~Lett.~{\bf 89},
128301 (2002). 

\bibitem{Bechinger17} 
M.~Brunner and C. Bechinger, Phys.~Rev.~Lett.~{\bf 88}, 248302 (2002);
K.~Mangold, P.~Leiderer, and C.~Bechinger, Phys.~Rev.~Lett.~{\bf 90}, 158302
(2003).

\bibitem{Curtis18}
J.E.~Curtis, B.A. Koss, and D.G. Grier, Opt. Comm.~{\bf 207}, 169 (2002). 

\bibitem{Viovy19}
W.D.~Volkmuth and R.H.~Austin, Nature (London) {\bf 358}, 600 (1992);
C.-F.~Chou, O. Bakajin, S.W.P. Turner, T.A.J. Duke, S.S. Chan,
E.C. Cox, H.G. Craighead, and R.H. Austin,  
Proc.~Natl.~Acad.~Sci. U.S.A. {\bf 96}, 13762 (1999); 
J.-L.~Vioy, Rev.~Mod.~Phys.~{\bf 72}, 813 (2000).

\bibitem{Weiss20}
D.~Weiss, M.L. Roukes, A. Menschig, P. Grambow, K. von Klitzing,
and G. Weimann, Phys.~Rev.~Lett.~{\bf 66}, 2790 (1991);
G.R.~Nash, S.J. Bending, M. Riek, and K. Eberl, 
Phys. Rev.~B {\bf 63}, 113316 (2001);
W.~Breuer, D.~Weiss, and V. Umansky, Physica E {\bf 12}, 216 (2002).

\bibitem{Grynberg21}
G.~Grynberg and C.~Robilliard, Phys. Rep. {\bf 5-6}, 335 (2001);
M.~Schiavoni, F.-R. Carminati, L. Sanchez-Palencia, F. Renzoni,
and G. Grynberg, Europhys.~Lett.~{\bf 59}, 493 (2002). 

\bibitem{Gronbech}
N.~Gr{\o}nbech-Jensen, Int.~J.~Mod.~Phys.~C {\bf 7}, 873 (1996).

\end{references}
\end{document}